\documentclass[10pt,letterpaper,twocolumn]{article}
\usepackage[latin9]{inputenc}
\usepackage{array}
\usepackage{rotating}
\usepackage{float}
\usepackage{multirow}
\usepackage{amsmath}
\usepackage{amssymb}
\usepackage{graphicx}
\usepackage{esint}
\usepackage[unicode=true]
 {hyperref}

\makeatletter

\pdfpageheight\paperheight
\pdfpagewidth\paperwidth

\providecommand{\tabularnewline}{\\}
\floatstyle{ruled}
\newfloat{algorithm}{tbp}{loa}
\providecommand{\algorithmname}{Algorithm}
\floatname{algorithm}{\protect\algorithmname}

\usepackage{icb}
\usepackage{times}
\usepackage{epsfig}
\usepackage{graphicx}

\usepackage{colortbl}
\usepackage[table]{xcolor}

\icbfinalcopy 

\ificbfinal\fi

\newcommand\blfootnote[1]{%
  \begingroup
  \renewcommand\thefootnote{}\footnote{#1}%
  \addtocounter{footnote}{-1}%
  \endgroup
}

\definecolor{darkgreen}{cmyk}{0.5, 0, 1, 0.5}
\definecolor{lightgray}{gray}{0.9}

\makeatother

\begin{document}

\title{Obfuscating Keystroke Time Intervals to Avoid Identification and
Impersonation}

\author{John~V.~Monaco\\
U.S.~Army Research Laboratory\\
Aberdeen, MD\\
\texttt{\small{}john.v.monaco2.civ@mail.mil} \and Charles~C.~Tappert\\
 Pace University\\
 Pleasantville, NY\\
 \texttt{\small{}ctappert@pace.edu}}

\maketitle
\thispagestyle{empty}

\blfootnote{This research was supported in part by an appointment to the Postgraduate Research Participation Program at the U.S.~Army Research Laboratory administered by the Oak Ridge Institute for Science and Education through an inter agency agreement between the U.S.~Department of Energy and USARL.}

\newcommand{\cg}{\cellcolor{lightgray}}
\begin{abstract}
There are numerous opportunities for adversaries to observe user behavior
remotely on the web. Additionally, keystroke biometric algorithms
have advanced to the point where user identification and soft biometric
trait recognition rates are commercially viable. This presents a privacy
concern because masking spatial information, such as IP address, is
not sufficient as users become more identifiable by their behavior.
In this work, the well-known Chaum mix is generalized to a scenario
in which users are separated by both space and time with the goal
of preventing an observing adversary from identifying or impersonating
the user. The criteria of a behavior obfuscation strategy are defined
and two strategies are introduced for obfuscating typing behavior.
Experimental results are obtained using publicly available keystroke
data for three different types of input, including short fixed-text,
long fixed-text, and long free-text. Identification accuracy is reduced
by 20\% with a 25 ms random keystroke delay not noticeable to the
user.
\end{abstract}

\section{Introduction}

The issue of privacy in behavior monitoring has recently started to
gain attention. It is not clear how many organizations routinely track
user behavior and to what extent this information is being used. On
the Internet, the situation is further complicated by the prevalence
of third-party content which allows identifying attributes and user
behavior to be seen by a number of websites as a result of visiting
only a single page \cite{mayer2012third}. Behavior tracking is motivated
by targeted advertising, analytics, and law enforcement. The laws
and regulations currently surrounding behavior tracking are relatively
lax, although a majority of users negatively view behavioral targeting.
In 2012, a study by Pew Research Group found 68\% of respondents to
be ``not okay'' with behavioral targeted advertising, and a 2010
USA Today/Gallup poll showed 67\% of respondents felt that behavioral
targeting should not be legal \cite{pew2012,gallup2010}. Behavior
tracking capabilities are extended when user keystrokes are added
to the mix \cite{chairunnanda2011privacy}.

The simplicity and ubiquitous nature of keystroke behavior makes it
attractive as a biometric modality for computer users. There are numerous
commercial applications and government interest in the area of keystroke
biometrics. To name a few, keystroke biometrics has recently been
considered as a means of offering students verified certificates for
completing a massively open online course (MOOC) \cite{maas2014offering}.
Funding through DARPA's Active Authentication program has helped the
field advance considerably, creating partnerships between government,
academia, and industry \cite{guidorizzi2013security}. One of the
most commercially successful keystroke biometric applications is delivered
by BehavioSec, a Swedish company that utilizes typing behavior, among
other factors, to verify the legitimacy of online transactions. In
2015, BehavioSec verified over 1.5B transactions from 15M users across
20 different banks \cite{behaviosec2015}.

Behavioral patterns and identifying attributes may be unintentionally
leaked through keystroke timing information, which can be remotely
observed without a victim's knowledge or consent. Masking spatial
information, such as IP address through TOR \cite{tor2015}, is futile
when temporal information, such as keystroke timings, can be used
for identification. The consequences of an adversary being able to
observe a victim's keystrokes are twofold. First, the victim may be
identified by their typing behavior. Even without having previously
observed the victim's typing behavior, identifying attributes such
as age, gender, handedness, and native language, can be resolved with
reasonable accuracy \cite{brizan2015utilizing,idrus2014soft}. Second,
the victim may be impersonated through a generative model of typing
behavior, potentially enabling an adversary to gain access to a system
that implements keystroke biometric access control.

This paper discusses the implications of typing behavior that is observable
by a third party over the web and proposes several mechanisms aimed
at preserving anonymity. The goals, constraints, and theoretical limits
of keystroke behavior obfuscation are explored. Section \ref{sec:Motivation}
provides further motivation and rationale for obfuscating typing behavior
and Section \ref{sec:background} reviews some background material
and related work. Two obfuscation strategies are introduced in Section
\ref{sec:Masking-strategies} followed by an empirical evaluation
using three types of keystroke input from publicly available databases
in Section \ref{sec:Case-study}. Finally, Section \ref{sec:Conclusions}
concludes the paper.

\section{Motivation\label{sec:Motivation}}

Humans generate events over a wide range of time scales separated
by orders of magnitude, as exemplified by Newell's time scale of human
action \cite{newell1994unified}. At the lowest level are biological
events, e.g., the firing of individual neurons on the order of microseconds,
and at the highest level are group and social dynamics which emerge
as a result of higher-frequency events. There is generally less privacy
as one moves up (down in frequency) in Newell's time scale. For example,
high frequency brain and heart activity, measured through electroencephalogram
(EEG) and electrocardiogram (ECG), require specialized sensors and
the presence of the user. On the other hand, lower frequency events
such as email, financial transactions, and mobile device interactions,
can typically be observed by a remote third party. Keystrokes fall
in the cognitive band, which lies between the biological and social
bands and where events occur on the order of milliseconds to seconds.
Since keystrokes are ubiquitous with modern computers and often contain
sensitive information, they are the subject of some privacy issues.

The standard desktop computer measures key-press and key-release events
with a clock precision of about 16 ms and resolution\footnote{Clock resolution is the degree to which a measurement can be made
and clock precision is the degree to which a measurement can be repeated.} of at least 1 ms \cite{killourhy2008effect}. Keystroke events are
typically recorded by a keylogger in either a system-wide context
via a hook registered in the kernel or in a sandbox environment, such
as within a single page of a web browser. In some cases, simply disabling
the logging software on the client is sufficient to ensure that a
third party cannot observe the victim's keystrokes. An exception to
this rule comes in the form of interactive applications, which generate
network traffic immediately after a key is pressed. Even if third
party logging software is disabled on the client, e.g., only trusted
Javascript sources are enabled, network timing information can be
used to build a profile of typing behavior since keystrokes result
in the immediate transmission of packets over a network. Key-press
times can be obtained remotely without installing a keylogger on the
victim's computer by observing the network traffic generated from
an interactive application, such as SSH in interactive mode \cite{song2001timing}
and Google Suggestions service \cite{tey2013keystroke}. A timing
attack on SSH network traffic timestamps collected during password
entry can provide about 1 bit of information in cracking a password.
\cite{song2001timing} verified that the key-press latencies can be
reliably determined from the packet inter-arrival times from interactive
SSH traffic, as the time between the actual press of a key and packet
creation by the kernel is negligible.

\subsection{Identification}

Identification is performed by enrolling a user's keystroke samples
into a database during a trusted session. The samples make up the
user's keystroke template. For fixed input, such as a password, this
may take the form of various key hold durations (the time from key
press to key release) and key press/key release latencies (the time
between successive key presses or key release to key press), such
as in \cite{killourhy2009comparing}. For free-text input, a set of
descriptive statistics on the various time intervals may be defined,
such as in \cite{tappert2009keystroke}. Note that in this work, only
timing information is considered. Some works that deal with free-text
input also consider linguistic features in an attempt to capture the
author's stylometry, or writing style. This work does not attempt
to mask information that may be leaked through linguistic analysis,
although there have been other attempts in successfully doing so \cite{kacmarcik2006obfuscating}.

The identification of soft biometric traits without a user's prior
enrollment is also a real possibility. In this scenario, a user's
typing behavior may reveal soft attributes such as age, gender, handedness,
and native language. Such an attack is feasible with several publicly
available databases that contain soft biometric labels and the relative
ease in collecting a new database with the desired labels through,
e.g., Amazon's Mechanical Turk. In \cite{idrus2014soft}, age, gender,
and handedness are classified with between 80\% and 100\% accuracy
using both fixed text and free text input. This database is made publicly
available. In \cite{brizan2015utilizing}, handedness, gender, and
native language are classified with accuracies that significantly
deviate from chance.

\subsection{Impersonation}

Impersonation is performed by mimicking keystroke behavior with the
intent of being recognized as the victim. This represents a non-zero-effort
attack on behalf of an impostor. With access to the victim's computer,
a spoofing attack such as in \cite{rahman2013snoop} can be utilized.
Observing a victim's keystrokes directly allows for typing behavior
to be easily replicated. Without any knowledge of the victim's typing
pattern, an attack such as \cite{serwadda2013examining} can be used.
The latter work used a template enumeration technique, in which an
independent keystroke database was used to reduce the search space
of typing behavior. This works well for shorter strings and assumes
that the verification system allows multiple attempts.

\cite{monaco2015spoofing} provides empirical evidence for a two-state
generative model of typing behavior in which the user can be in either
a passive or active state. Given key-press latencies with missing
key names, the model is then used to predict the key-press latencies
of a user by exploiting the linear relationship between inter-key
distance and key-press latency. The proposed generative model uses
this partial information to perform a key-press-only sample-level
attack on a victim's keystroke dynamics template. Results show that
some users are more susceptible to this type of attack than others.
For about 10\% of users, the spoofed samples obtain classifier output
scores of at least 50\% of those obtained by authentic samples, and
with at least 50 observed keystrokes the chance of successful verification
over a zero-effort attack doubles on average.

\section{Background\label{sec:background}}

\begin{center}
\begin{figure}
\begin{centering}
\includegraphics[width=0.9\columnwidth]{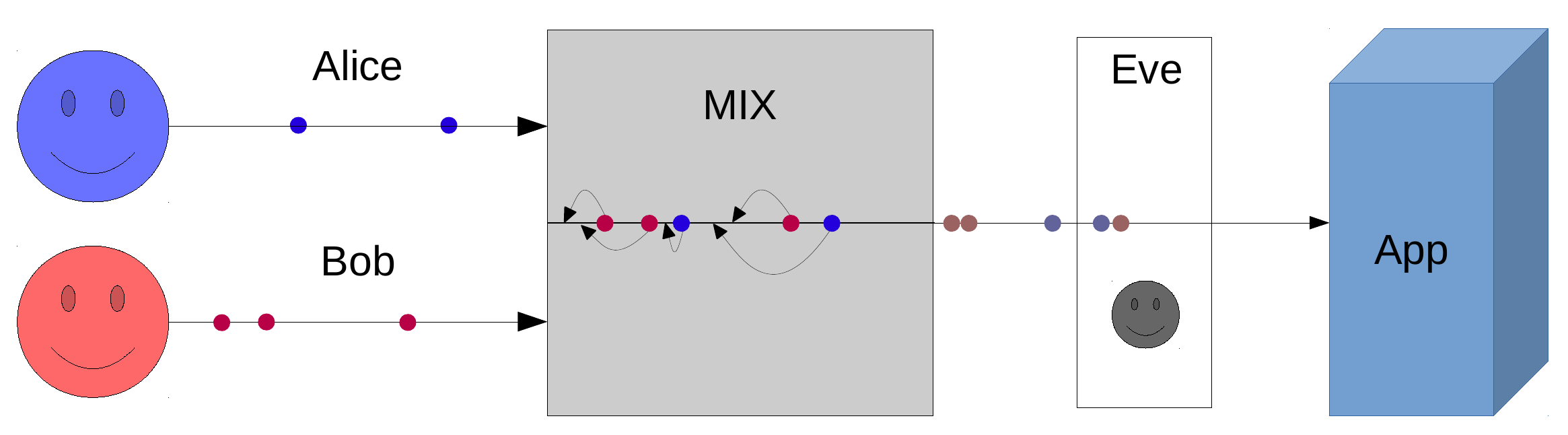}
\par\end{centering}

\caption{Chaum mix. Eve is able to observe a single arrival process and must
discern which user owns each event.\label{fig:Chaum-mix}}
\end{figure}

\par\end{center}

There are primarily two goals in obfuscating keystroke behavior. The
first is to limit an adversary's capability to identify a user. Under
this goal, the user wishes to remain anonymous within a pool of $U$
users. Anonymity is achieved when the probability of correctly identifying
the user is no greater than $\frac{1}{U}$. With cooperation from
all the users in the pool, it is relatively easy to obtain perfect
anonymity for every user. Each user needs only to behave in some predefined
way agreed upon by the pool. If the behavior of every user is exactly
the same, the best strategy for identification is simply a random
guess out of the $U$ users.

The second goal is to limit an adversary's capability to predict user
behavior, which is necessary for impersonation. This goal is quite
different, in that the user wishes for the time and duration of a
future keystroke to be unpredictable after having generated $N$ keystrokes.
Perfect unpredictability is achieved when the expected symmetric mean
absolute percentage error (SMAPE) of the predictions go to 1 \cite{raghavan2013hidden}.
The SMAPE can go to 1 when either the actual time interval between
events, $\tau_{N+1}$, or predicted time interval between events,
$\hat{\tau}_{N+1}$, is infinite. The actual time interval is infinite
when the user plans on waiting indefinitely before generating another
keystroke. The predicted time interval is infinite if the adversary's
model predicts an infinite estimated delay until the next keystroke.
Neither of these situations are practical and thus perfect unpredictability
can generally not be achieved.

Both goals in obfuscating keystroke behavior call for different strategies.
Consider two extreme examples. In the first, every user decides to
generate keystrokes with exactly the same frequency. That is, the
time between events $\tau$ is the same for every user. Under this
strategy, perfect anonymity is achieved. However, each user's actions
can also be predicted exactly and the masking strategy fails to address
the predictability criterion. Consider a different strategy, in which
each user generates keystrokes according to a Poisson process. Let
the users have rates $\lambda_{1}\ll\dots\ll\lambda_{U}$. On a per-user
basis, predictive accuracy is relatively low, as the time intervals
are independent and identically distributed, following an exponential
distribution. However, users can be easily identified over time by
their expected time interval, or event frequency.

A more effective strategy would be to let each user generate events
according to a Poisson process with rate $\lambda$. With cooperation
from every user, perfect anonymity is achieved as the Poisson process
is memoryless and the resulting distributions of time intervals from
each user will appear to be the same. Unfortunately, this strategy
is generally not practical since it requires the cooperation from
every user. Some users who naturally type at a rate much lower than
$\lambda$ would be forced to increase their speed. Conversely, users
who naturally type at a rate much higher than $\lambda$ would be
forced to slow down.

Chaum mixes can be used to provide anonymity to users behind a router,
in which cooperation is assumed \cite{chaum1981untraceable}. This
scenario is shown in Figure \ref{fig:Chaum-mix}, where users Alice
and Bob are behind a router and wish to transmit a series of packets.
The \emph{mix} reorders the packets by introducing a random delay
to each packet. An eavesdropper, Eve, observes the sequence of reordered
packets before they reach the application, however cannot discern
whether each packet came from Alice or Bob. The identity and temporal
behavior of Alice and Bob are protected by the mix.

\begin{center}
\begin{figure}
\begin{centering}
\includegraphics[width=0.9\columnwidth]{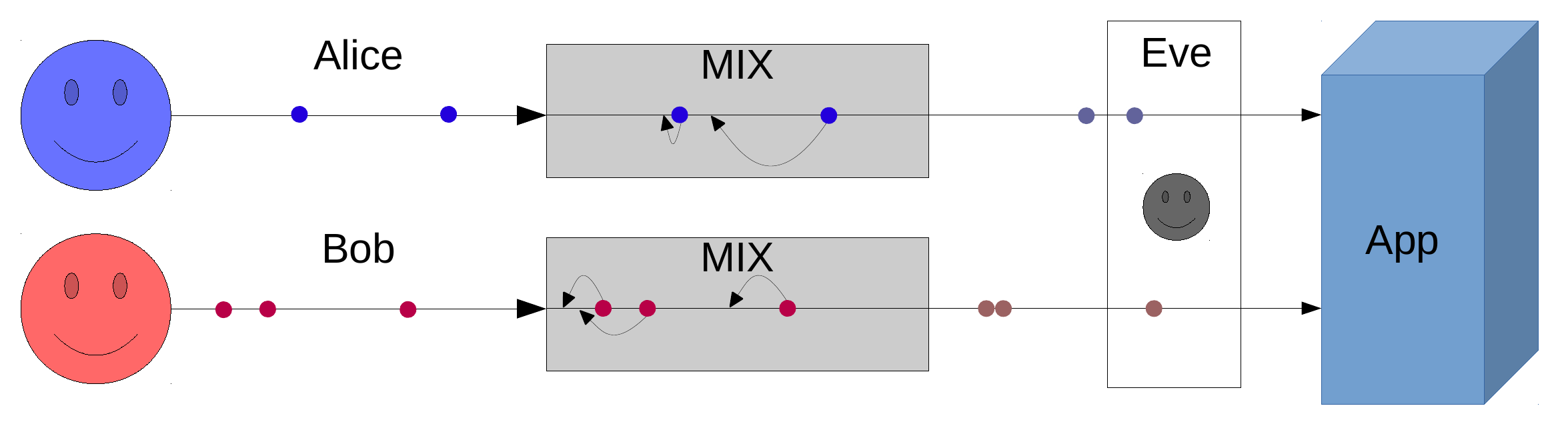}
\par\end{centering}

\caption{User mix. Eve is able to observe multiple arrival processes and must
discern which user owns each process.\label{fig:user-mix}}
\end{figure}

\par\end{center}

The anonymity of the mixing strategy is given by the expected entropy
of the observed packets, 
\begin{equation}
\mathcal{A}=\lim_{N\rightarrow\infty}\frac{1}{N}\mathsf{E}\left[H\left(U_{1},\dots,U_{N}\right)\right]\label{eq:anonymity}
\end{equation}
where $U_{n}$ is the identity of the $n^{\mathrm{th}}$ observed
packet and $H$ is the entropy function. Without any constraints,
the optimal strategy that maximizes anonymity can be shown to require
an infinite delay \cite{mishra2011anonymity}. This is due to the
mix waiting for all packets from both Alice and Bob and then randomly
selecting one of the ${2N \choose N}$ permutations for the packet
reordering. This strategy is not practical for a number of reasons,
one being that it would require an infinite buffer size. With a finite
buffer size, the upper bound on the maximum-achievable anonymity decreases.
Despite this, a reasonable level of anonymity can be achieved with
practical constraints \cite{venkitasubramaniam2008anonymity}.

The scenario considered in this work is a generalization of the Chaum
mix, shown in Figure \ref{fig:user-mix}. Since the Chaum mix requires
the cooperation of all the users, i.e., every user must be on the
same network, it is a global obfuscation strategy, or \emph{global
mix}. If the assumption of cooperation is relaxed, then an obfuscation
strategy without cooperation must be developed. This type of strategy
is provided by a \emph{user mix}. A user mix is appropriate when there
is no cooperation between users, or the events from different users
are separated by both space and time\footnote{For example, consider keystrokes recorded by a web application. The
keystrokes from two different sessions could have been recorded days
apart and come from entirely different network locations.}.

Compared to the Chaum mix that operates globally on events from every
user, the user mix operates exclusively on events from only one user.
The mix for each user is atomic, in the sense that it operates independently
of every other user. Different users may employ different obfuscation
strategies or no strategy at all. The mix separates the \emph{generating
process}, the sequence of events from the perspective of the user,
from the \emph{arrival process}, the sequence of events as seen by
the application. The goal of the mix is to provide anonymity and unpredictability
at the arrival process.

This scenario is depicted in Figure \ref{fig:user-mix}, where Alice
and Bob both generate events and wish to remain anonymous and unpredictable.
An eavesdropper, Eve, is able to observe the sequence of events from
each user after passing through the mix, and she knows that there
are two users in the system. From the perspective of Eve, it is not
clear which sequence belongs to each user, as she can only observe
the arrival processes. The technical conditions of the user mix are
summarized as follows.

\begin{tabular*}{1\columnwidth}{@{\extracolsep{\fill}}c>{\raggedright}p{0.9\columnwidth}}
1. & {\small{}Alice and Bob have zero knowledge of each other's behavior
and don't cooperate.}\tabularnewline
2. & {\small{}Eve can see multiple arrival processes, but cannot discern
which process belongs to each user}\tabularnewline
3. & {\small{}Each mix has a source of randomness unknown to Eve.}\tabularnewline
4. & {\small{}Event order is preserved by the mix.}\tabularnewline
\end{tabular*}

\section{Obfuscation strategies\label{sec:Masking-strategies}}

\begin{table}
\begin{centering}
\renewcommand{\arraystretch}{0.9}%
\begin{tabular}{|>{\raggedright}p{0.22\columnwidth}|>{\raggedright}p{0.68\columnwidth}|}
\hline 
\textbf{\small{}Finite} & {\small{}The expected delay between the user and the arrival process
should not grow unbounded.}\tabularnewline
\hline 
\textbf{\small{}Anonymous} & {\small{}The mix should make it difficult to identify the user.}\tabularnewline
\hline 
\textbf{\small{}Unpredictable} & {\small{}The mix should make it difficult to predict future behavior.}\tabularnewline
\hline 
\end{tabular}
\par\end{centering}

\caption{Desirable properties of a user mix.\label{tab:mix-properties}}
\end{table}

The user mix delays events by temporarily storing them in a buffer
before releasing them to the application. Without access to future
events, and under the constraint that events cannot be permuted, the
only operation that can be performed is to delay an event. An event
consists of a key press or key release. The constraint that events
cannot be permuted ensures that the characters appear at the arrival
process in the order and form (e.g., case or special symbol) they
were generated.

There are a few caveats in obfuscating temporal behavior. Since the
only action is to delay, a lag is introduced between the generation
of an event by the user and the observation of the event in the application.
Thus, it may not be possible to use a mix in some real-time systems.
For many human-computer interaction applications, a small lag is acceptable
and would not be noticed by the user. As the lag increases, the movement
time and error rate on behalf of the user also increase \cite{mackenzie1993lag}.

A user mix should possess several properties, summarized in Table
\ref{tab:mix-properties}. The expected lag between the generated
events and observed events should be finite. The size of the event
buffer ultimately determines how large the lag can grow to. If the
tolerated lag is unbounded, then the number of events that need to
be stored will eventually exceed the size of the buffer. A user mix
should also protect the user from being identified out of a population
of users. This condition is more difficult to satisfy, as it requires
at least some global knowledge of other users in the system. A mix
that provides anonymity will emulate the temporal behavior of the
``typical'' user. Finally, a mix should make it difficult to reproduce
and predict the temporal behavior of a user.

Analogous to the tradeoff between type I and type II errors on the
ROC curve, there is a direct tradeoff between time lag and the obfuscation
capability of the mix. As the lag decreases, the dependence between
the arrival process and the generating process increase. As the lag
increases, the potential to reduce dependence increases. It is desirable
to have a mix that maximally reduces the dependence between the generating
process and the arrival process for a given lag.

\begin{figure}
\begin{centering}
\renewcommand{\arraystretch}{0.9}%
\begin{tabular}{|>{\centering}p{0.15\columnwidth}|>{\raggedright}p{0.75\columnwidth}|}
\hline 
{\footnotesize{}Symbol} & {\footnotesize{}Description}\tabularnewline
\hline 
\hline 
{\footnotesize{}$t_{n}$} & {\footnotesize{}time of events at the arrival process}\tabularnewline
\hline 
{\footnotesize{}$t_{n}^{\circ}$} & {\footnotesize{}time of events at the generating process}\tabularnewline
\hline 
{\footnotesize{}$\tau_{n}$} & {\footnotesize{}time interval between events at the arrival process}\tabularnewline
\hline 
{\footnotesize{}$\tau_{n}^{\circ}$} & {\footnotesize{}time interval between events at the generating process}\tabularnewline
\hline 
{\footnotesize{}$\delta_{n}$} & {\footnotesize{}time lag between the generating and arrival processes}\tabularnewline
\hline 
\end{tabular}
\par\end{centering}

\begin{centering}
\includegraphics[width=1\columnwidth]{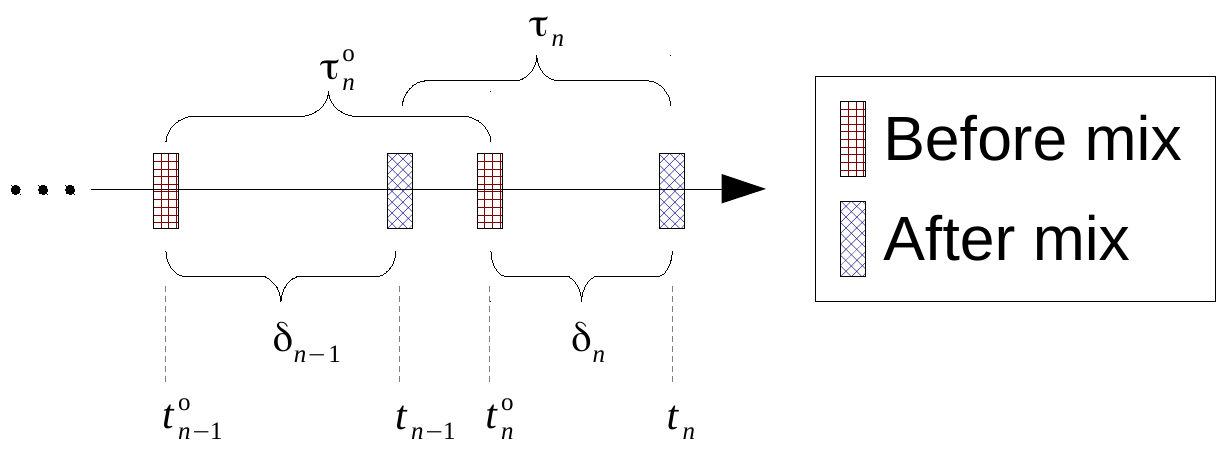}
\par\end{centering}

\caption{Summary of user mix operation and variables.\label{fig:Mix-variables}}
\end{figure}

Let $t_{n}^{\circ}$ and $\tau_{n}^{\circ}$ be the true time and
time interval of the $n^{\mathrm{th}}$ event generated by the user,
and $t_{n}$ and $\tau_{n}$ be the observed time and time interval
of the $n^{\mathrm{th}}$ event after passing through the mix. The
sequence of $t_{n}$, or alternatively $\tau_{n}$, constitute the
arrival process. The lag, or delay, between the generated event and
observed event is given by $\delta_{n}$. Since event order must be
preserved by the mix, it is necessary that $t_{n-1}+\delta_{n-1}<t_{n}+\delta_{n}$.
Figure \ref{fig:Mix-variables} summarize the variables of the user
mix.

Predictability can be measured by the dependence between $\tau^{\circ}$
and $\tau$. Anonymity is given by the dependence between $\tau^{A}$
and $\tau^{B}$ from two random instantiations of the mix, $A$ and
$B$. Dependence is measured by the mutual information between two
continuous random variables, given by
\begin{equation}
I\left(X,Y\right)=\int_{Y}\int_{X}f\left(x,y\right)\log\left[\frac{f\left(x,y\right)}{f\left(x\right)f\left(y\right)}\right]dxdy\label{eq:mi}
\end{equation}
where $f\left(x,y\right)$ is the joint probability density function
of $x$ and $y$.

A mix with that provides perfect unpredictability has $I\left(\tau^{\circ},\tau\right)=0$.
Perfect anonymity is achieved when $I\left(\tau^{A},\tau^{B}\right)=0$.
Both can be achieved with an unlimited buffer size and delay. To demonstrate
this, consider a mix that samples $\tau$ from a uniform distribution,
i.e., $\tau\sim\mathcal{U}\left(0,k\right)$. Let $\tau_{n}$ be the
time interval of the arrival process. If $\tau_{n+1}^{\circ}<\tau_{n}$,
then the lag between the user and the arrival process must increase
by at least $\tau_{n}-\tau_{n+1}^{\circ}$ and as $n\rightarrow\infty$,
the lag increases unbounded. In this case, $I\left(\tau^{\circ},\tau\right)=0$
since $f\left(\tau^{\circ},\tau\right)=f\left(\tau^{\circ}\right)f\left(\tau\right)$,
i.e., the observed time intervals are independent from the actual
time intervals. If $\tau_{n+1}^{\circ}>\tau_{n}$, then the time interval
on the arrival process is no longer independent from the actual time
intervals since $\tau_{n+1}>\tau_{n+1}^{\circ}-\tau_{n}$. The next
sample will have $\tau_{n+1}\sim\mathcal{U}\left(\tau_{n+1}^{\circ}-\tau_{n},\tau_{n+1}^{\circ}-\tau_{n}+1\right)$,
or alternatively $P\left(\tau_{n+1}=s\right)=\frac{1}{k}$ where $\tau-\tau_{n}<s<\tau_{n+1}^{\circ}-\tau_{n}+k$.
Consequently, the dependence guarantees that $f\left(\tau^{\circ},\tau\right)>f\left(\tau^{\circ}\right)f\left(\tau\right)$
and $I\left(\tau^{\circ},\tau\right)>0$. Therefore, in practice it
is necessary to introduce a lag constraint in minimizing the dependence
between $\tau^{\circ}$ and $\tau$.

There are primarily two ways a time interval mix can add noise to
the time intervals of the arrival process. Let $\delta$ be the lag
between $\tau^{\circ}$ and $\tau$. The first is by generating a
random delay between the actual events that occur at time $t^{\circ}$
resulting in the time $t^{\circ}+\delta$ at the arrival process.
This type of mix is a \emph{delay mix}. The second way is to generate
$\tau$ directly, referred to as an \emph{interval mix}.

\subsection{Delay mix}

The delay mix introduces noise to the time intervals of the arrival
process by randomly delaying each event. For the $n^{\mathrm{th}}$
event at time $t_{n}^{\circ}$, the delay mix generates a random delay
$\delta_{n}$ to produce the event time at the arrival process, $t_{n}=t_{n}^{\circ}+\delta_{n}$.
Time intervals of the arrival process are given by $\tau_{n}=\left(t_{n}^{\circ}+\delta_{n}\right)-\left(t_{n-1}^{\circ}+\delta_{n-1}\right)$.
A uniform distribution can be used to generate delays. Let $\delta_{n}\sim\mathcal{U}\left(\max\left(\delta_{n-1}-\tau_{n}^{\circ},0\right),\Delta\right)$,
where $\Delta$ is the upper bound on the delay. The lower bound of
$\max\left(\delta_{n-1}-\tau_{n}^{\circ},0\right)$ is necessary to
ensure the event order is preserved. Without this constraint, events
will become permuted if $\tau_{n}^{\circ}+\delta_{n}<\delta_{n-1}$.
The process is summarized in Algorithm \ref{alg:delay-mix}.

\begin{algorithm}
\begin{enumerate}
\item \textbf{Initialize}: Let $\tau_{0}^{\circ}=\infty$ and $t_{0}^{\circ}=0$
\item \textbf{Generate}: Let $l_{n}=\max\left(\delta_{n-1}-\tau_{n}^{\circ},0\right)$
be the lower bound of the $n^{\mathrm{th}}$ delay and $\delta_{n}\sim\mathcal{U}\left(l,\Delta\right)$
be a random delay. The event time on the arrival process is given
by $t_{n}=t_{n}^{\circ}+\delta_{n}$
\end{enumerate}
\caption[Delay mix]{Delay mix.\label{alg:delay-mix}}
\end{algorithm}

The expected delay is bounded above by $\Delta$. To see this, consider
the two scenarios: $\tau_{n}^{\circ}<\delta_{n-1}$ and $\tau_{n}^{\circ}>\delta_{n-1}$.
If $\tau_{n}^{\circ}<\delta_{i-1}$, then the expected lag is simply
$\mathsf{E}\left[\delta_{n}\right]=\frac{\Delta}{2}$ since the lower
bound will always be 0. If $\tau_{n}^{\circ}>\delta_{n-1}$, then
$\mathsf{E}\left[\delta_{i}\right]=\frac{\Delta+\delta_{n-1}-\tau_{n}^{\circ}}{2}$
which is bounded above by $\Delta$. Given $\tau_{0}=\infty$, it
is true that $0<\delta_{1}<\Delta$. By induction, it is also true
that $\delta_{n-1}<\Delta$ since $\tau_{n}^{\circ}>0$ .

\subsubsection{Example}

As an example, consider events occurring at times $t^{\circ}=\left[0,5,7,11,14\right]$
passing through a delay mix with $\Delta=7$. The actual time intervals
are $\tau^{\circ}=\left[\infty,5,2,4,3\right]$. The trace of each
variable is shown in Table \ref{tab:Delay-mix-example}.

\begin{table}[h]
\begin{centering}
\renewcommand{\arraystretch}{0.8}%
\begin{tabular}{|c|c|c|c|c|c|c|}
\hline 
{\footnotesize{}Event} & {\footnotesize{}$t^{\circ}$} & {\footnotesize{}$\tau^{\circ}$} & {\footnotesize{}$l$} & {\footnotesize{}$\delta$} & {\footnotesize{}$t$} & {\footnotesize{}$\tau$}\tabularnewline
\hline 
\hline 
{\footnotesize{}0} & {\footnotesize{}0} & {\footnotesize{}$\infty$} & {\footnotesize{}0} & {\footnotesize{}3} & {\footnotesize{}3} & {\footnotesize{}$\infty$}\tabularnewline
\hline 
{\footnotesize{}\cg1} & {\footnotesize{}\cg5} & {\footnotesize{}\cg5} & {\footnotesize{}\cg0} & {\footnotesize{}\cg6} & {\footnotesize{}\cg11} & {\footnotesize{}\cg8}\tabularnewline
\hline 
{\footnotesize{}2} & {\footnotesize{}7} & {\footnotesize{}2} & {\footnotesize{}4} & {\footnotesize{}5} & {\footnotesize{}12} & {\footnotesize{}1}\tabularnewline
\hline 
{\footnotesize{}\cg3} & {\footnotesize{}\cg11} & {\footnotesize{}\cg4} & {\footnotesize{}\cg1} & {\footnotesize{}\cg5} & {\footnotesize{}\cg16} & {\footnotesize{}\cg4}\tabularnewline
\hline 
{\footnotesize{}4} & {\footnotesize{}14} & {\footnotesize{}3} & {\footnotesize{}6} & {\footnotesize{}6} & {\footnotesize{}20} & {\footnotesize{}4}\tabularnewline
\hline 
\end{tabular}
\par\end{centering}

\caption{Delay mix example where $t^{\circ}$ and $\tau^{\circ}$ are the actual
time and time interval of the $n^{\mathrm{th}}$ event, $l$ is the
lower bound of the delay, $\delta$ is the random delay, and $t$
and $\tau$ are the time and time interval of the arrival process.\label{tab:Delay-mix-example}}
\end{table}

\subsection{Interval mix}

Instead of generating delays, the time intervals of the arrival process
can be modeled explicitly. The goal of the mix is to minimize the
dependence between $\tau^{\circ}$ and $\tau$, and this strategy
will give the mix greater control over the resulting time intervals.
While the delay mix generates $\delta_{n}$ at each time step to produce
arrival process time $t_{n}=t_{n}^{\circ}+\delta_{n}$, the interval
mix generates $\tau_{n}$ to produce $t_{n}=t_{n-1}+\tau_{n}$. With
an infinite delay, this allows the $\tau_{n}$ to be generated independently
of $\tau_{n}^{\circ}$. With a finite delay, the $\tau_{n}$ are constrained
by the event rate of the user.

At each time step $t_{n}$, let the desired time interval be $\dot{\tau}_{n}\sim\mathcal{U}\left(0,u_{n-1}\right)$
where $u_{n-1}$ is an upper bound parameter. If $t_{n-1}+\dot{\tau_{n}}>t_{n}^{\circ},$
then the $n^{\mathrm{th}}$ event is delayed by $\delta_{n}=t_{n-1}+\dot{\tau_{n}}-t_{n}^{\circ}$
to get the arrival process time $t_{n}=t_{n-1}+\dot{\tau}_{n}$. If
$t_{n-1}+\dot{\tau_{n}}\le t_{n}^{\circ}$, then $\delta_{n}=0$ and
the event is released immediately. The upper bound $u$ is updated
as $u_{n}=\max\left\{ u_{n-1}+b\left[t_{n}-\left(t_{n-1}+\dot{\tau}_{n}\right)\right],\epsilon\right\} $,
where $b$ is a parameter that controls the rate at which $u_{n}$
can change. This update moves the expected time interval in the direction
of the rate of the user. This process is summarized in Algorithm \ref{alg:Interval-mix}.

\begin{algorithm}
\begin{enumerate}
\item \textbf{Initialize}: Let $t_{0}^{\circ}=t_{0}=0$, $\tau_{0}^{\circ}=\tau_{0}=\infty$,
and $u_{1}>0$.
\item \textbf{Generate}: Let $\dot{\tau}_{n}\sim\mathcal{U}\left(0,u_{n}\right)$
be the desired time interval and $\dot{t}_{n}=t_{n-1}+\dot{\tau_{n}}$
be the desired time of the arrival process. The arrival process time
is given by $t_{n}=\max\left(\dot{t}_{n},t_{n}^{\circ}\right)$.
\item \textbf{Update}: Let $u_{n+1}=\max\left[u_{n}+b\left(t_{n}^{\circ}-\dot{t}_{n}\right),\epsilon\right]$
\end{enumerate}
\caption[Interval mix]{Interval mix.\label{alg:Interval-mix}}
\end{algorithm}

\subsubsection{Example}

An example is used to demonstrate the interval mix with parameter
$b=1$. Similar as before, the actual event times are given by $t^{\circ}=\left[0,5,7,11,14\right]$
with time intervals $\tau^{\circ}=\left[\infty,5,2,4,3\right]$. At
the first event, there is no delay between the actual event time and
the arrival process, therefore $t_{0}=t_{0}^{\circ}=0$. The parameter
updates and time intervals of the arrival process are shown in Table
\ref{tab:Interval-mix-example}. For simplicity, values are sampled
from a discrete uniform distribution. The starting value of $u$ is
chosen to be $7$ to demonstrate the adaptability of the mix to the
user's behavior in several iterations. The starting value of $u$
can be any positive value and will quickly converge to an appropriate
range.

\begin{table}
\begin{centering}
\renewcommand{\arraystretch}{0.8}%
\begin{tabular}{|c|c|c|c|c|c|c|c|c|}
\hline 
{\footnotesize{}Event} & {\footnotesize{}$t^{\circ}$} & {\footnotesize{}$\tau^{\circ}$} & {\footnotesize{}$u$} & {\footnotesize{}$\dot{t}$} & {\footnotesize{}$\dot{\tau}$} & {\footnotesize{}$\delta$} & {\footnotesize{}$t$} & {\footnotesize{}$\tau$}\tabularnewline
\hline 
\hline 
{\footnotesize{}0} & {\footnotesize{}0} & {\footnotesize{}$\infty$} & {\footnotesize{}-} & {\footnotesize{}-} & {\footnotesize{}-} & {\footnotesize{}0} & {\footnotesize{}0} & {\footnotesize{}$\infty$}\tabularnewline
\hline 
{\footnotesize{}\cg1} & {\footnotesize{}\cg5} & {\footnotesize{}\cg5} & {\footnotesize{}\cg7} & {\footnotesize{}\cg3} & {\footnotesize{}\cg3} & {\footnotesize{}\cg0} & {\footnotesize{}\cg5} & {\footnotesize{}\cg5}\tabularnewline
\hline 
{\footnotesize{}2} & {\footnotesize{}7} & {\footnotesize{}2} & {\footnotesize{}9} & {\footnotesize{}11} & {\footnotesize{}6} & {\footnotesize{}4} & {\footnotesize{}11} & {\footnotesize{}6}\tabularnewline
\hline 
{\footnotesize{}\cg3} & {\footnotesize{}\cg11} & {\footnotesize{}\cg4} & {\footnotesize{}\cg5} & {\footnotesize{}\cg15} & {\footnotesize{}\cg4} & {\footnotesize{}\cg4} & {\footnotesize{}\cg15} & {\footnotesize{}\cg4}\tabularnewline
\hline 
{\footnotesize{}4} & {\footnotesize{}14} & {\footnotesize{}3} & {\footnotesize{}1} & {\footnotesize{}16} & {\footnotesize{}1} & {\footnotesize{}2} & {\footnotesize{}16} & {\footnotesize{}1}\tabularnewline
\hline 
\end{tabular}
\par\end{centering}

\caption{Interval mix example where $t^{\circ}$is the actual event time, $\tau^{\circ}$
is the actual time interval, $u$ is the interval distribution parameter,
$\dot{t}$ is the desired event time, $t$ is the arrival process
event time, and $\tau$ is the arrival process time interval.\label{tab:Interval-mix-example}}
\end{table}

\section{Case study\label{sec:Case-study}}

The two time interval obfuscation strategies introduced in Section
\ref{sec:Masking-strategies} are empirically evaluated using publicly
available keystroke datasets with previously published results. Three
different types of keystroke input are considered: short fixed-text
(e.g., password or PIN), long fixed-text (e.g., copying or transcribing
several sentences), and long free-text (e.g., response to an open-ended
question).

Each dataset contains labeled attributes for identity, age, handedness,
and gender. The short fixed-text comes from \cite{idrus2014soft},
the long fixed-text from \cite{bello2010collection}, and the long
free-text from \cite{monaco2013recent}. The short fixed-text dataset
contains 110 users with 50 samples per user. Each sample contains
one entry in which the user typed a short passphrase ranging from
17 to 24 characters as described in \cite{idrus2014soft}. The long
fixed-text dataset is a combination of the fixed text from \cite{bello2010collection}
and \cite{monaco2013recent}, during which users copied short sentences
and fables. The combined dataset contains 101 users with 10 samples
per user and $123\pm38$ characters per sample. The free text dataset
is also from \cite{monaco2013recent} and contains 127 users with
10 samples per user. Users were required to respond to open-ended
and essay style questions. The long responses were sliced to create
samples of length $123\pm38$ to match the distribution of the long
fixed-text dataset.

Classification of each target variable is performed by a random forest
classifier with features and fallback hierarchy as described in \cite{monaco2013recent}.
The RandomForest classifier in the \texttt{sklearn} Python package
is utilized with 200 estimators and the same features for all three
types of keystroke input. Classification accuracy (ACC) is obtained
by a stratified 10 fold cross validation for the identity target.
For the age, handedness, and gender targets, a 110, 101, and 127 cross-fold
validation is used for the short fixed-text, long fixed-text, and
long free-text, respectively. In each fold, a single user's samples
are used as the testing set and the remainder of the population as
the training set. This emulates an open system, in which soft attributes
for the target user are classified without having previously observed
that user's keystrokes. The sizes of each class in the training sets
are not altered. Thus, the baseline prediction rates coincide with
the proportion of the largest class, which are 0.51 (age $\ge$ 30),
0.68 (male), 0.88 (right handed) for age, gender, and handedness targets
for all three datasets combined. Predictions are made for both the
press-press latency and duration using the mean time interval up to
the observed event, $\widehat{PP}_{n}=<PP_{1}^{n-1}>$ and $\widehat{DU}_{n}=<DU_{1}^{n-1}>$
where $PP_{n}^ {}$ and $DU_{n}$ are the press-press latency and
duration of the $n^{\mathrm{th}}$ keystroke, respectively. Source
code for the experiments in this work is available at \href{https://github.com/vmonaco/keystroke-obfuscation}{https://github.com/vmonaco/keystroke-obfuscation}.

\begin{figure}
\begin{centering}
\includegraphics[width=1\columnwidth]{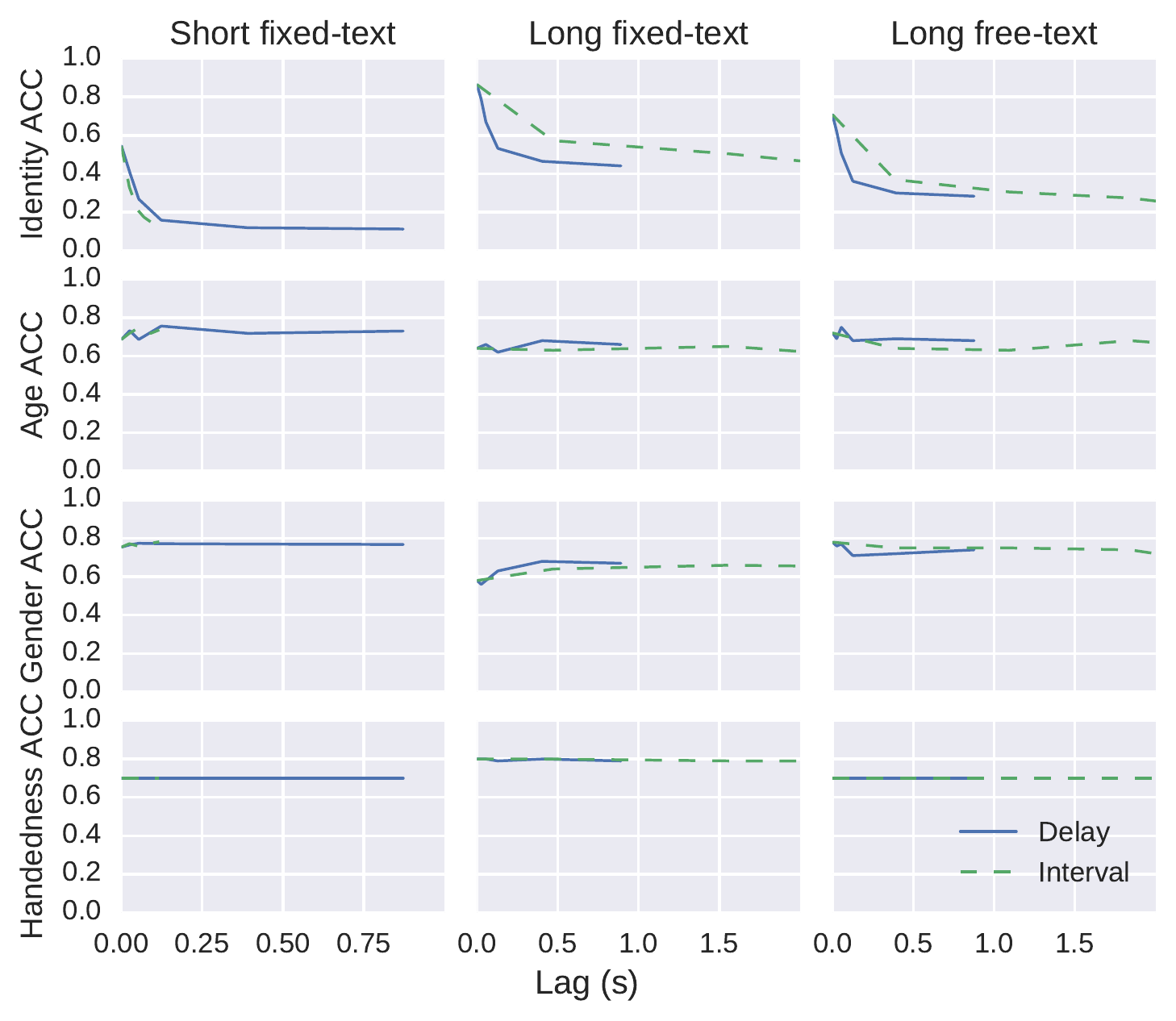}
\par\end{centering}

\caption{Time lag vs classification accuracy for each type of keystroke input
and target variable.\label{fig:Time-lag-vs-acc}}
\end{figure}

Using each dataset, classification accuracies for each target variable
are obtained before and after applying each masking strategy. The
mean lag $\bar{\delta}$ between the generating process and arrival
process is also calculated for each parameter choice. Table \ref{tab:delay-mix-performance}
contains experimental results using the delay mix for various values
of $\Delta$ and Table \ref{tab:Interval-mix-performance} contains
experimental results using the interval mix for various values of
$b$. The classification accuracies of each mix as a function of the
mean lag $\bar{\delta}$ are summarized in Figure \ref{fig:Time-lag-vs-acc}.

\begin{table}
\begin{centering}
\renewcommand{\arraystretch}{0.8}%
\begin{tabular}{|c||r|r|r|r|r|r|r|r|}
\cline{2-9} 
\multicolumn{1}{c|}{} & {\footnotesize{}$\Delta$} & {\footnotesize{}$\bar{\delta}$} & {\footnotesize{}Id.} & {\footnotesize{}Age} & {\footnotesize{}Gen.} & {\footnotesize{}Han.} & {\footnotesize{}PP} & {\footnotesize{}DU}\tabularnewline
\hline 
\multirow{6}{*}{\begin{turn}{90}
{\footnotesize{}Short fixed-text}
\end{turn}} & {\footnotesize{}0} & {\footnotesize{}0} & {\footnotesize{}0.55} & {\footnotesize{}0.69} & {\footnotesize{}0.75} & {\footnotesize{}0.70} & {\footnotesize{}0.20} & {\footnotesize{}0.14}\tabularnewline
\cline{2-9} 
 & \cg{\footnotesize{}50} & \cg{\footnotesize{}26} & \cg{\footnotesize{}0.41} & \cg{\footnotesize{}0.73} & \cg{\footnotesize{}0.77} & \cg{\footnotesize{}0.70} & \cg{\footnotesize{}0.21} & \cg{\footnotesize{}0.18}\tabularnewline
\cline{2-9} 
 & {\footnotesize{}100} & {\footnotesize{}54} & {\footnotesize{}0.27} & {\footnotesize{}0.69} & {\footnotesize{}0.77} & {\footnotesize{}0.70} & {\footnotesize{}0.22} & {\footnotesize{}0.26}\tabularnewline
\cline{2-9} 
 & \cg{\footnotesize{}200} & \cg{\footnotesize{}123} & \cg{\footnotesize{}0.16} & \cg{\footnotesize{}0.76} & \cg{\footnotesize{}0.77} & \cg{\footnotesize{}0.70} & \cg{\footnotesize{}0.25} & \cg{\footnotesize{}0.32}\tabularnewline
\cline{2-9} 
 & {\footnotesize{}500} & {\footnotesize{}391} & {\footnotesize{}0.12} & {\footnotesize{}0.72} & {\footnotesize{}0.77} & {\footnotesize{}0.70} & {\footnotesize{}0.27} & {\footnotesize{}0.35}\tabularnewline
\cline{2-9} 
 & \cg{\footnotesize{}1000} & \cg{\footnotesize{}872} & \cg{\footnotesize{}0.11} & \cg{\footnotesize{}0.73} & \cg{\footnotesize{}0.77} & \cg{\footnotesize{}0.70} & \cg{\footnotesize{}0.30} & \cg{\footnotesize{}0.38}\tabularnewline
\hline 
\hline 
\multirow{6}{*}{\begin{turn}{90}
{\footnotesize{}Long fixed-text}
\end{turn}} & {\footnotesize{}0} & {\footnotesize{}0} & {\footnotesize{}0.86} & {\footnotesize{}0.64} & {\footnotesize{}0.58} & {\footnotesize{}0.80} & {\footnotesize{}0.31} & {\footnotesize{}0.16}\tabularnewline
\cline{2-9} 
 & \cg{\footnotesize{}50} & \cg{\footnotesize{}26} & \cg{\footnotesize{}0.79} & \cg{\footnotesize{}0.65} & \cg{\footnotesize{}0.56} & \cg{\footnotesize{}0.80} & \cg{\footnotesize{}0.31} & \cg{\footnotesize{}0.19}\tabularnewline
\cline{2-9} 
 & {\footnotesize{}100} & {\footnotesize{}55} & {\footnotesize{}0.67} & {\footnotesize{}0.66} & {\footnotesize{}0.58} & {\footnotesize{}0.80} & {\footnotesize{}0.32} & {\footnotesize{}0.26}\tabularnewline
\cline{2-9} 
 & \cg{\footnotesize{}200} & \cg{\footnotesize{}129} & \cg{\footnotesize{}0.53} & \cg{\footnotesize{}0.62} & \cg{\footnotesize{}0.63} & \cg{\footnotesize{}0.79} & \cg{\footnotesize{}0.34} & \cg{\footnotesize{}0.31}\tabularnewline
\cline{2-9} 
 & {\footnotesize{}500} & {\footnotesize{}404} & {\footnotesize{}0.46} & {\footnotesize{}0.68} & {\footnotesize{}0.68} & {\footnotesize{}0.80} & {\footnotesize{}0.34} & {\footnotesize{}0.34}\tabularnewline
\cline{2-9} 
 & \cg{\footnotesize{}1000} & \cg{\footnotesize{}891} & \cg{\footnotesize{}0.44} & \cg{\footnotesize{}0.66} & \cg{\footnotesize{}0.67} & \cg{\footnotesize{}0.79} & \cg{\footnotesize{}0.34} & \cg{\footnotesize{}0.35}\tabularnewline
\hline 
\hline 
\multirow{6}{*}{\begin{turn}{90}
{\footnotesize{}Long free-text}
\end{turn}} & {\footnotesize{}0} & {\footnotesize{}0} & {\footnotesize{}0.71} & {\footnotesize{}0.72} & {\footnotesize{}0.78} & {\footnotesize{}0.70} & {\footnotesize{}0.30} & {\footnotesize{}0.15}\tabularnewline
\cline{2-9} 
 & \cg{\footnotesize{}50} & \cg{\footnotesize{}26} & \cg{\footnotesize{}0.62} & \cg{\footnotesize{}0.69} & \cg{\footnotesize{}0.76} & \cg{\footnotesize{}0.70} & \cg{\footnotesize{}0.31} & \cg{\footnotesize{}0.18}\tabularnewline
\cline{2-9} 
 & {\footnotesize{}100} & {\footnotesize{}54} & {\footnotesize{}0.51} & {\footnotesize{}0.75} & {\footnotesize{}0.77} & {\footnotesize{}0.70} & {\footnotesize{}0.32} & {\footnotesize{}0.24}\tabularnewline
\cline{2-9} 
 & \cg{\footnotesize{}200} & \cg{\footnotesize{}126} & \cg{\footnotesize{}0.36} & \cg{\footnotesize{}0.68} & \cg{\footnotesize{}0.71} & \cg{\footnotesize{}0.70} & \cg{\footnotesize{}0.33} & \cg{\footnotesize{}0.30}\tabularnewline
\cline{2-9} 
 & {\footnotesize{}500} & {\footnotesize{}394} & {\footnotesize{}0.30} & {\footnotesize{}0.69} & {\footnotesize{}0.72} & {\footnotesize{}0.70} & {\footnotesize{}0.33} & {\footnotesize{}0.33}\tabularnewline
\cline{2-9} 
 & \cg{\footnotesize{}1000} & \cg{\footnotesize{}876} & \cg{\footnotesize{}0.28} & \cg{\footnotesize{}0.68} & \cg{\footnotesize{}0.74} & \cg{\footnotesize{}0.70} & \cg{\footnotesize{}0.34} & \cg{\footnotesize{}0.35}\tabularnewline
\hline 
\end{tabular}
\par\end{centering}

\caption{Delay mix experimental results. Id.=identity, Gen.=gender, Han.=handedness,
PP=press-press latency SMAPE, DU=duration SMAPE.\label{tab:delay-mix-performance}}
\end{table}

\begin{table}
\begin{centering}
\renewcommand{\arraystretch}{0.8}%
\begin{tabular}{|c||r@{\extracolsep{0pt}.}l|r|r|r|r|r|r|r|}
\cline{2-10} 
\multicolumn{1}{c|}{} & \multicolumn{2}{c|}{{\footnotesize{}$b$}} & {\footnotesize{} $\bar{\delta}$} & {\footnotesize{}Id.} & {\footnotesize{}Age} & {\footnotesize{}Gen.} & {\footnotesize{}Han.} & {\footnotesize{}PP} & {\footnotesize{}DU}\tabularnewline
\hline 
\multirow{6}{*}{\begin{turn}{90}
{\footnotesize{}Short fixed-text}
\end{turn}} & {\footnotesize{}0}&{\footnotesize{}0} & {\footnotesize{}0} & {\footnotesize{}0.55} & {\footnotesize{}0.69} & {\footnotesize{}0.75} & {\footnotesize{}0.70} & {\footnotesize{}0.20} & {\footnotesize{}0.14}\tabularnewline
\cline{2-10} 
 & \multicolumn{2}{c|}{\cg{\footnotesize{}0.1}} & \cg{\footnotesize{}24} & \cg{\footnotesize{}0.33} & \cg{\footnotesize{}0.72} & \cg{\footnotesize{}0.77} & \cg{\footnotesize{}0.70} & \cg{\footnotesize{}0.19} & \cg{\footnotesize{}0.18}\tabularnewline
\cline{2-10} 
 & {\footnotesize{}0}&{\footnotesize{}5} & {\footnotesize{}50} & {\footnotesize{}0.21} & {\footnotesize{}0.74} & {\footnotesize{}0.76} & {\footnotesize{}0.70} & {\footnotesize{}0.24} & {\footnotesize{}0.27}\tabularnewline
\cline{2-10} 
 & \multicolumn{2}{c|}{\cg{\footnotesize{}1.0}} & \cg{\footnotesize{}70} & \cg{\footnotesize{}0.17} & \cg{\footnotesize{}0.75} & \cg{\footnotesize{}0.78} & \cg{\footnotesize{}0.70} & \cg{\footnotesize{}0.29} & \cg{\footnotesize{}0.36}\tabularnewline
\cline{2-10} 
 & {\footnotesize{}1}&{\footnotesize{}5} & {\footnotesize{}90} & {\footnotesize{}0.15} & {\footnotesize{}0.72} & {\footnotesize{}0.77} & {\footnotesize{}0.70} & {\footnotesize{}0.33} & {\footnotesize{}0.43}\tabularnewline
\cline{2-10} 
 & \multicolumn{2}{c|}{\cg{\footnotesize{}2.0}} & \cg{\footnotesize{}117} & \cg{\footnotesize{}0.13} & \cg{\footnotesize{}0.74} & \cg{\footnotesize{}0.78} & \cg{\footnotesize{}0.70} & \cg{\footnotesize{}0.37} & \cg{\footnotesize{}0.49}\tabularnewline
\hline 
\hline 
\multirow{6}{*}{\begin{turn}{90}
{\footnotesize{}Long fixed-text}
\end{turn}} & {\footnotesize{}0}&{\footnotesize{}0} & {\footnotesize{}0.00} & {\footnotesize{}0.86} & {\footnotesize{}0.64} & {\footnotesize{}0.58} & {\footnotesize{}0.80} & {\footnotesize{}0.31} & {\footnotesize{}0.16}\tabularnewline
\cline{2-10} 
 & \multicolumn{2}{c|}{\cg{\footnotesize{}0.1}} & \cg{\footnotesize{}471} & \cg{\footnotesize{}0.57} & \cg{\footnotesize{}0.63} & \cg{\footnotesize{}0.64} & \cg{\footnotesize{}0.80} & \cg{\footnotesize{}0.32} & \cg{\footnotesize{}0.26}\tabularnewline
\cline{2-10} 
 & {\footnotesize{}0}&{\footnotesize{}5} & {\footnotesize{}1566} & {\footnotesize{}0.50} & {\footnotesize{}0.65} & {\footnotesize{}0.66} & {\footnotesize{}0.79} & {\footnotesize{}0.40} & {\footnotesize{}0.37}\tabularnewline
\cline{2-10} 
 & \multicolumn{2}{c|}{\cg{\footnotesize{}1.0}} & \cg{\footnotesize{}2544} & \cg{\footnotesize{}0.42} & \cg{\footnotesize{}0.59} & \cg{\footnotesize{}0.65} & \cg{\footnotesize{}0.79} & \cg{\footnotesize{}0.46} & \cg{\footnotesize{}0.46}\tabularnewline
\cline{2-10} 
 & {\footnotesize{}1}&{\footnotesize{}5} & {\footnotesize{}3570} & {\footnotesize{}0.35} & {\footnotesize{}0.64} & {\footnotesize{}0.66} & {\footnotesize{}0.80} & {\footnotesize{}0.50} & {\footnotesize{}0.53}\tabularnewline
\cline{2-10} 
 & \multicolumn{2}{c|}{\cg{\footnotesize{}2.0}} & \cg{\footnotesize{}4218} & \cg{\footnotesize{}0.32} & \cg{\footnotesize{}0.65} & \cg{\footnotesize{}0.59} & \cg{\footnotesize{}0.80} & \cg{\footnotesize{}0.54} & \cg{\footnotesize{}0.58}\tabularnewline
\hline 
\hline 
\multirow{6}{*}{\begin{turn}{90}
{\footnotesize{}Long free-text}
\end{turn}} & {\footnotesize{}0}&{\footnotesize{}0} & {\footnotesize{}0} & {\footnotesize{}0.71} & {\footnotesize{}0.72} & {\footnotesize{}0.78} & {\footnotesize{}0.70} & {\footnotesize{}0.30} & {\footnotesize{}0.15}\tabularnewline
\cline{2-10} 
 & \multicolumn{2}{c|}{\cg{\footnotesize{}0.1}} & \cg{\footnotesize{}385} & \cg{\footnotesize{}0.37} & \cg{\footnotesize{}0.64} & \cg{\footnotesize{}0.75} & \cg{\footnotesize{}0.70} & \cg{\footnotesize{}0.32} & \cg{\footnotesize{}0.26}\tabularnewline
\cline{2-10} 
 & {\footnotesize{}0}&{\footnotesize{}5} & {\footnotesize{}1093} & {\footnotesize{}0.30} & {\footnotesize{}0.63} & {\footnotesize{}0.75} & {\footnotesize{}0.70} & {\footnotesize{}0.40} & {\footnotesize{}0.39}\tabularnewline
\cline{2-10} 
 & \multicolumn{2}{c|}{\cg{\footnotesize{}1.0}} & \cg{\footnotesize{}1840} & \cg{\footnotesize{}0.27} & \cg{\footnotesize{}0.68} & \cg{\footnotesize{}0.74} & \cg{\footnotesize{}0.70} & \cg{\footnotesize{}0.47} & \cg{\footnotesize{}0.49}\tabularnewline
\cline{2-10} 
 & {\footnotesize{}1}&{\footnotesize{}5} & {\footnotesize{}2321} & {\footnotesize{}0.23} & {\footnotesize{}0.65} & {\footnotesize{}0.68} & {\footnotesize{}0.70} & {\footnotesize{}0.51} & {\footnotesize{}0.56}\tabularnewline
\cline{2-10} 
 & \multicolumn{2}{c|}{\cg{\footnotesize{}2.0}} & \cg{\footnotesize{}3783} & \cg{\footnotesize{}0.18} & \cg{\footnotesize{}0.64} & \cg{\footnotesize{}0.77} & \cg{\footnotesize{}0.70} & \cg{\footnotesize{}0.56} & \cg{\footnotesize{}0.61}\tabularnewline
\hline 
\end{tabular}
\par\end{centering}

\caption{Interval mix experimental results.\label{tab:Interval-mix-performance}}
\end{table}

\section{Conclusions\label{sec:Conclusions}}

The results in this work suggest that it is possible to obfuscate
keystroke behavior with a delay that is not noticeable to the user,
however the time lag between the user and the application remains
a practical constraint. With a 25 ms delay, identification accuracy
is reduced by approximately 20\% on average, and in most cases a 500
ms delay is needed to halve the identification accuracy. Soft biometric
trait classification accuracies, which are initially near chance accuracy,
are relativity unaffected by the obfuscated keystrokes. Results may
differ in a scenario where, e.g., only the unlabeled testing data
is obfuscated. Prediction errors of the duration and press-press latency
increase as expected using both types of obfuscation strategy.

An important point is that the proposed methods operate at the keystroke
event level and not at the network packet level. Only in certain situations
do keystroke events correspond to network packets, such as interactive
web applications. Another realistic scenario is one in which Eve resides
on a workstation, and Alice and Bob are users typing on a peripheral
USB keyboard. The proposed obfuscation strategies could be implemented
in hardware, e.g., as a device that sits between the keyboard and
the computer. This device would ensure a given level of anonymity
by randomly delaying the USB events before they reach the workstation.
The hardware implementation may prove more secure than the software
implementation, and it avoids interference with the application (such
as disabling Javascript on a web page and subsequently losing other
functionality). 

Future work should investigate additional obfuscation strategies and
scenarios, such as training on normal data and testing obfuscated
data. Additionally, a major limitation of this work is that the proposed
obfuscation strategies ignore the possibility that only a single user
utilizes a mix and can be easily detected. The susceptibility of additional
features and classifiers should also be investigated with the proposed
obfuscation strategies. Temporal behavior obfuscation is a relatively
unexplored area. The described strategies can be applied to network
traffic to hinder device fingerprinting and network traffic classification
based on packet inter arrival times, such as those described in \cite{radhakrishnan2014gtid}.
It can also be applied to lower-frequency events, such as financial
transactions, although it is not clear if there would be any benefit
in doing so.

{\small{}\bibliographystyle{ieee}
\bibliography{icb2016,monaco}
}{\small \par}
\end{document}